\documentclass[twocolumn,prb,aps,showpacs]{revtex4}
\usepackage{epsfig,latexsym,graphicx}
\def \ba {\begin{eqnarray}}
\def \ea {\end{eqnarray}}
\def \vk {{\bf k}}

\def \ra {\rightarrow}

\def \bi {Bi$_2$Sr$_2$CaCu$_2$O$_{8+\delta}$ }
\def \bsl {Bi$_2$Sr$_{2-x}$La$_x$CuO$_6$ }

\begin{document}

\title{The dynamically induced Fermi arcs and Fermi pockets in two dimensions:
 a model for underdoped cuprates }

\author{Han-Yong Choi\email[To whom the
correspondences should be addressed:~]{ hychoi@skku.edu}}
\affiliation{Department of Physics and Institute for Basic
Science Research, SungKyunKwan University, Suwon 440-746, Korea.
\\ School of Physics, Korea Institute for Advanced Study, Seoul 130-722, Korea}

\author{Seung Hwan Hong}
\affiliation{Department of Physics and Institute for Basic
Science Research, SungKyunKwan University, Suwon 440-746, Korea.}

\begin{abstract}

We investigate the effects of the dynamic bosonic fluctuations on
the Fermi surface reconstruction in two dimensions as a model for
the underdoped cuprates. At energies larger than the boson energy
$\omega_b$, the dynamic nature of the fluctuations is not
important and the quasi-particle dispersion exhibits the shadow
feature like that induced by a static long range order. At lower
energies, however, the shadow feature is pushed away by the finite
$\omega_b$. The detailed low energy features are determined by the
bare dispersion and the coupling of quasi-particles to the dynamic
fluctuations. We present how these factors reconstruct the Fermi
surface to produce the Fermi arcs or the Fermi pockets, or their
coexistence. Our principal result is that the dynamic nature of
the fluctuations, without invoking a yet-to-be-established
translational symmetry breaking hidden order, can produce the
Fermi pocket centered away from the $(\pi/2,\pi/2)$ towards the
zone center which may coexist with the Fermi arcs. This is
discussed in comparison with the experimental observations.

\end{abstract}



\maketitle

\section{Introduction}

The ``Fermi arc'' picture was advanced by the angle-resolved
photo-emission spectroscopy (ARPES) to understand the enigmatic
pseudo-gap state in the underdoped
cuprates.\cite{Marshall96prl,Norman98nature,Shen05science,Kanigel06naturephys}
The ARPES, with its momentum resolution capability, established
that in this pseudo-gap state the gapped region is mainly in the
$(0,\pi)$ and $(\pi,0)$ region while the Fermi surface (FS)
exists in the diagonal direction. Then, the picturesque view of
the pseudo-gap state is that the gapless portion of the FS forms
an open ended arc, rather than a closed loop as in ordinary
metals. It is extremely difficult to understand the abrupt
truncation of the FS in the Brillouin zone. The Fermi arc has
thus puzzled the physics community and triggered enormous
research efforts.\cite{Timusk99rpp}

This Fermi arc picture was challenged by the observations of the
quantum oscillation under the applied magnetic field
$H$.\cite{LeBoeuf07nature,Sebastian08nature,Doiron-Leyraud07nature}
The transport and thermodynamic properties exhibit the periodic
oscillations as a function of the inverse magnetic field. The
standard interpretation is in terms of the closed loop of the FS,
or, the Fermi pockets. The oscillation is due to the quantizated
Landau levels, and its periodicity is proportional to the area of
the Fermi pocket. It is found to be only a few percent of the FS
area of optimally or overdoped cuprates. In the theory of usual
metals, such a small FS would require a change of translational
symmetry from overdoped to underdoped cuprates. The problem is
that there is no direct experimental evidence for the
translational symmetry breaking for the compounds exhibiting the
small FS. Moreover, the Fermi pocket is at odds with the Fermi
arc picture from ARPES. Although the ARPES were done above $T_c$
with no magnetic field and the quantum oscillations in the low
$T$ and strong external field, the views they advance, the Fermi
arc and Fermi pocket, seem contradictory each other and need to
be reexamined.

The recent laser ARPES on the single layer \bsl compounds by Meng
$et~al.$ \cite{Meng09nature} is indeed very interesting in this
regard. They observed, with the improved resolution, that the
ungapped portion of FS forms a closed loop, $e.g.$ the Fermi
pocket, rather than the Fermi arcs at the doping levels of 11 and
12 \% for \bsl. Moreover, the center of the Fermi pocket is
shifted from the $(\pi/2,\pi/2)$ toward the zone center ($\Gamma$
point). The translational symmetry breaking, let alone its
yet-to-be-established existence, can not explain their results
because a salient feature of the reconstructed FS induced by the
broken translational symmetry of period doubling is that the FS
is symmetric with respect to the $(\pi,0)-(0,\pi)$ line.

Here, we wish to understand the Fermi pocket centered away from
the $(\pi/2,\pi/2)$ point without invoking the translation
symmetry breaking in terms of the $dynamic$ bosonic fluctuations.
We first consider a dynamical collective mode coupled with
quasiparticles at the antiferromagnetic (AF) wave vectors only
(the correlation length $\xi\ra\infty$) for simplicity and
illustration of basic ideas. Then, the more realistic cases of
finite $\xi$ are presented with self-consistent numerical
calculations.

There have been many attempts to understand the Fermi arcs and
pockets in the cuprates. Each of them has discrepancies with the
experimental observations such as the shape, location, or the
spectral
weight.\cite{Wen96prl,Ng05prb,Yang06prb,Norman07prb,Sachdev09prb,Greco09prl}
On the other hand, the dynamic nature of the bosonic fluctuatoins
peaked at $(\pi,\pi)$, without invoking a hidden order which
breaks the translational symmetry, can produce the FS evolution
from the large FS to Fermi arc to Fermi pocket as the coupling is
increased. More specifically, it can induce (1) the Fermi pocket
centered away from the $(\pi/2,\pi/2)$ towards the $\Gamma$
point, (2) the ratio of the spectral weight at the back side of
the Fermi pocket to the inner side is about $10^{-2}$, (3)
coexistence of the Fermi pocket and the large main FS, and (4)
the dispersion kink along the nodal direction at energy $\approx
0.05$ eV. These are in agreement with the recent laser ARPES
experiment of Meng $et~al$ \cite{Meng09nature} and numerous
previous experimental
reports.\cite{Damascelli03rmp,Bogdanov00prl,Kaminski00prl,Lanzara01nature}

After the bare band dispersion is determined there are three
factors which affect the Fermi surface reconstructions: the
fluctuations correlation length $\xi$, coupling constant $\alpha$,
and the boson frequency scale $\omega_b$. More discussion about
their possible microscopic origin and relation will made later in
Sec.\ V in connection with other approaches. For now, we first
take the Einstein mode of $\omega_b$ for simplicity.
$\omega_b=0.05$ eV was chosen to match the kink
energy.\cite{Bogdanov00prl,Kaminski00prl,Lanzara01nature} We will
also consider the realistic frequency dependent bosonic spectrum
recently deduced by Bok $et~al$\cite{Bok10prb} by inverting the
laser ARPES on \bi. We then perform detailed numerical
calculations and show that the dynamic nature of the collective
mode can account for the FS evolution without introducing a
yet-to-be-established hidden order parameter.

\section{Idea and Formulation}

We consider the renormalization of the fermions due to the
coupling to the dynamic bosonic fluctuations $F({\bf q},\omega)$
with the coupling vertex $\alpha(\vk,\vk')$. The self-energy of
the fermion is given by\cite{Kampf90prb}
 \ba
 \label{self-energy}
\Sigma(\vk,\omega)= \int_{-\infty}^{\infty}d\epsilon
 \int_{-\infty}^{\infty}d\epsilon'
 \frac{f(\epsilon)+n(-\epsilon')}{\epsilon+\epsilon'-\omega-i\delta}
 \nonumber \\
\times \sum_{\vk'} A(\vk',\epsilon) \alpha^2
F(\vk,\vk',\epsilon'),
 \ea
where $A$ is the spectral function of the fermion, and $f$ and
$n$ are the Fermi and Bose distribution functions, respectively.
 \ba
 \label{spectral}
A(\vk',\epsilon)=-\frac1\pi Im \frac{1}{\epsilon-\xi_{\vk'}-\Sigma(\vk',\epsilon)}, \\
\alpha^2 F(\vk,\vk',\epsilon')= -\frac1\pi \alpha(\vk,\vk')^2 Im
V(\vk-\vk',\epsilon').
 \ea
We took the fluctuation spectrum of the following factorized
form:\cite{Kampf90prb}
 \ba
\alpha^2 F(\vk,\vk',\epsilon') = \alpha(\vk,\vk')^2 F(\epsilon')
\nonumber \\
 \times \sum_{{\bf Q} ={\pm\pi/a,\pm\pi/a}}
\frac{\Gamma/\pi}{(q_x-Q_x)^2+\Gamma^2}
\frac{\Gamma/\pi}{(q_y-Q_y)^2+\Gamma^2},
 \label{a2f}
 \ea
where $a$ is the lattice constant, ${\bf q}={\vk}'-{\vk}$, and
$\Gamma=\pi/\xi$. The coupling $\alpha$ may depend on the wave
vectors $\vk$ and $\vk'$, but for simplicity we will consider a
constant $\alpha$ and the Einstein model of frequency $\omega_b$
first.
 \ba
\alpha^2 F(\epsilon')= \alpha^2
 \left[ \delta(\epsilon'-\omega_b)- \delta(\epsilon'+\omega_b)
\right].
 \ea
Some remarks will be made on the more realistic frequency
dependence of $F({\bf q},\epsilon')$ and the momentum dependence
of $\alpha(\vk,\vk')$ later. Eqs.\ (\ref{self-energy}) and
(\ref{spectral}) constitute the coupled self-consistency
equations. They are solved self-consistently for the self-energy
via numerical iterations. A very similar problem was investigated
by Grilli $et~al.$ for the one dimensional electronic
systems.\cite{Grilli09prb} It is extended to two dimensions in
the present work fully self-consistently.

Let us first consider the simple case of $T\ra 0$ and $\Gamma\ra
0$ to gain underlying physics. That is, the boson mode is of a
delta function in both the energy and momentum channels.  Then, in
the limit $T\rightarrow 0$, Eq.\ (\ref{self-energy}) is reduced to
 \ba
 \label{self-consis0}
\Sigma(\vk,\omega)= -\alpha^2 \int_{-\infty}^{\infty} d\epsilon
\left[ \frac{\Theta(\epsilon)}{\epsilon+\omega_b-\omega-i\delta}
\right. \nonumber \\
 \left. + \frac{ \Theta(-\epsilon)}{\epsilon-\omega_b-\omega-i\delta}
\right]A({\bf k}_Q,\epsilon), \\
 Im \Sigma(\vk,\omega)= - \pi \alpha^2 \left[\Theta(\omega -\omega_b)
 A({\bf k}_Q,\omega-\omega_b) \right. \nonumber \\
\left. +\Theta(-\omega -\omega_b) A({\bf k}_Q,\omega+\omega_b)
\right],
 \ea
where ${\bf k}_Q=\vk +{\bf Q}$ and $\Theta$ is the step function.
A useful approximation is to take
 \ba
 \label{spectral1}
A(\vk,\epsilon)= \delta(\epsilon-\xi_{k}).
 \ea
We then have
 \ba
\Sigma(\vk,\omega) = \alpha^2 \left[ \frac{\Theta(\xi_{k_Q})} {
{\omega+i\delta- \xi_{k_Q}-\omega_b} } \right. \nonumber \\
 \left. +\frac{\Theta(-\xi_{k_Q})}
{ {\omega+i\delta-\xi_{k_Q}+\omega_b} } \right]
 =\frac{\alpha^2 }{\omega+i\delta- \tilde\xi_{k_Q}} ,
 \ea
with the definition
 \ba
\tilde\xi_{k_Q} = \xi_{k_Q}+sgn(\xi_{k_Q})\omega_b.
\label{tildexikq}
 \ea
The Green's function of quasi-particle (qp) is given by
 \ba
G(\vk,\omega)=\frac{1}{\omega-\xi_k -\frac{\alpha^2}{\omega-
\tilde\xi_{k_Q}}}. \label{green0}
 \ea
This form of the Green's function appeared previously in the
context of the pseudogap.\cite{Yang06prb,Sachdev09prb} The
coupling vertex $\alpha(\vk,\vk')$ of present approach corresponds
to the pseudogap $\Delta_R$ of Ref.\ \cite{Yang06prb}. It will be
interesting to check to what extent this mapping is valid. An
important distinction of the present approach is that the dynamics
of the bosonic fluctuations is explicitly built in via
$\tilde\xi_{k_Q} $ of Eq.\ (\ref{tildexikq}). It is precisely
this dynamics which gives rise to the Fermi arcs as we will see
now.

The qp dispersion $E(\vk)$ is determined by
 \ba
G^{-1}(\vk,\omega)=\omega-\xi_k-\Sigma(\vk,\omega)=0,
 \ea
which gives
 \ba
E_\pm(\vk) = \frac12 \left[\xi_k+\tilde\xi_{k_Q} \pm
\sqrt{(\xi_k-\tilde\xi_{k_Q})^2 +4\alpha^2} \right].
 \label{dispersion}
 \ea
The results may approximately be extended to the case of finite
correlation length $1/\xi \ne 0$ following Ref.\
\cite{Kuchinskii08jetp} by replacing the imaginary part of the
frequency by $ \delta=\hbar v_F/\xi$.

The Green's function may be cast into the form
 \ba
 \label{greenfunc}
G(\vk,\omega) = \frac{u_k^2}{\omega+i \delta-E_+}
+\frac{v_k^2}{\omega+i \delta-E_-},
 \ea
where the coherence factors are given by
 \ba
 \label{cohfactor}
u_k^2 = \frac12\left[1 +
\frac{\xi_k-\tilde\xi_{k_Q}}{\sqrt{(\xi_k-\tilde\xi_{k_Q})^2
+4\alpha^2}} \right],\nonumber \\
 v_k^2 = \frac12\left[1 -
\frac{\xi_k-\tilde\xi_{k_Q}}{\sqrt{(\xi_k-\tilde\xi_{k_Q})^2
+4\alpha^2}} \right] .
 \ea
The $E_+$ and $E_-$ represent, respectively, the electron and
hole bands. The spectral function $A(\vk,\omega)$ is then
 \ba
A(\vk,\omega) &=& -\frac1\pi Im G(\vk,\omega) \nonumber \\
 &=& u_k^2 \delta(\omega-E_+(\vk)) +v_k^2
\delta(\omega-E_-(\vk)).
 \label{specfun}
 \ea
The spectral function is directly probed by the ARPES.

\section{Preliminary analysis}

Before showing the detailed numerical results, we will first
present the preliminary analysis to gain underlying physics of
the problem. The bare dispersion of \bsl is taken as
 \ba
 \label{bband}
\xi_k &=& -2t[\cos(k_x a)+\cos(k_y a)] +4t' \cos(k_x a)\cos(k_y a)
\nonumber \\
 &-& 2t'' [\cos(2k_x a)+\cos(2k_y a)] -\mu,
 \ea
where $t=0.25,~t'=0.058,~t''=t'/2$ eV.\cite{hashimoto08prb} The FS
corresponding to the $\xi_k$ and $E_\pm(\vk)$ with $\alpha=0.1$
eV, $\omega_b=0.05$ eV, and $\mu=-0.208$ eV corresponding to the
slight underdoping of 12 \% are shown in Fig.\ 1. The nodal cut
of $k_x=k_y$ and several cuts parallel to it are also shown with
dashed lines.

\begin{figure}
\vspace{-0.cm}\label{fig1}
\includegraphics[width=6cm]{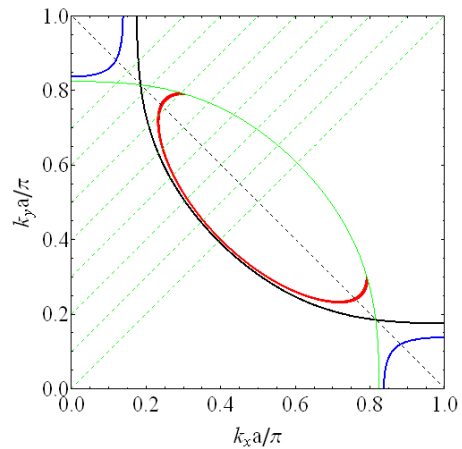}
 \vspace{-0.cm}
\caption{The reconstructed FSs for $\omega_b=0.05$, $\alpha=0.1$,
and $\mu=-0.208$ eV corresponding to $x=12$ \% doping. The thick
black curve is the bare FS determined by $\xi_k=0$ of Eq.\
(\ref{bband}) and the green curve is the shadow FS of
$\xi_{k+Q}=0$. The red curve around the $(\pi/2,\pi/2)$ point
shows the hole FS of $E_-(\vk)=0$. Notice that the outer portion
is gapped and the hole FS forms the Fermi arc. The blue curves
around $(0,\pi)$ and $(\pi,0)$ are the electron FS of
$E_+(\vk)=0$. As $\alpha$ increases it may disappear as shown in
Fig.\ 3. The nodal cut of $k_y=k_x$ and the parallel cuts of
$k_y=k_x+0.1 n$ with $n=1-7$ are indicated with dashed green
lines. The band dispersions along these cuts are shown in Fig.\
2. }
 \label{fig:FS}
\end{figure}

Along the cuts the qp dispersions are presented in Fig.\ 2 to
better reveal the dynamically generated gap close to the shadow
FS. Fig.\ 2(a) is the hole band dispersion $E_-(\vk)$ in solid
blue and electron band $E_+(\vk)$ in dashed green lines along the
nodal cut given by Eq.\ (\ref{dispersion}). The important point
is that the hole band dispersion exhibits the abrupt jump at $k_x
a/\pi \approx 0.6$, or the gap of about 2$\omega_b$. The
dynamically induced gap was noticed by Grilli $et~al.$ for the
one dimensional electronic systems.\cite{Grilli09prb}

The gap of $2\omega_b$ means that for $\omega=0$ there exists only
a single $\vk$ point which satisfies $\omega=E_-(\vk)$, while for
$|\omega| \gtrsim\omega_b$ there exist two $\vk$ points, one
close to the original FS and the other to the shadow FS. That is,
the shadow feature is present for $|\omega| \gtrsim\omega_b$, but
is absent for $\omega=0$. This is in accord with general
expectations: A physical system may have long (but finite) ranged
order-parameter spatial correlations which fluctuate with the
frequency $\omega_b$. The system then appears to be ordered above
$\omega_b$. For energies larger than $\omega_b$ with respect to
the Fermi energy the spectra should resemble an ordered system.
On the other hand, at lower energies electrons ``sense'' averaged
order-parameter fluctuations, and the system appears to be not
disturbed much from the one without the collective mode.

The blowup of the hole band dispersions is shown in Fig.\ 2(b)
along the cuts parallel to the nodal cut. Notice that the gap
survives beyond $k_y a/\pi=k_x a/\pi+0.4$. It simply means that
the gapless portion of the FS forms an open ended arc as shown
with the thick red solid curve in Fig.\ 1. {\it We stress that the
abrupt truncation of the FS, which seemed so puzzling, is
naturally understood in terms of the dynamic boson mode.}

\begin{figure}
\includegraphics[width=6cm]{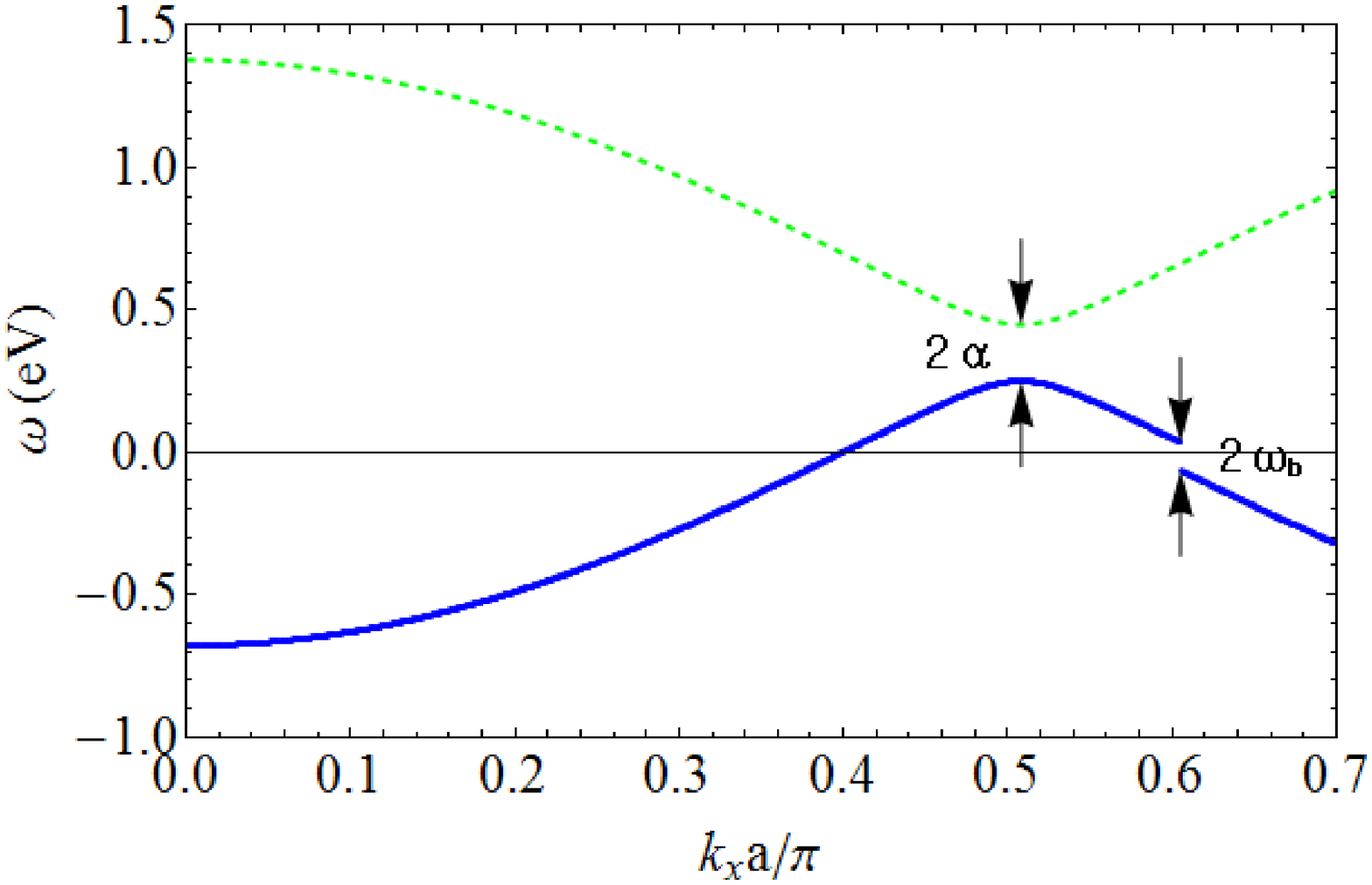}
\includegraphics[width=6cm]{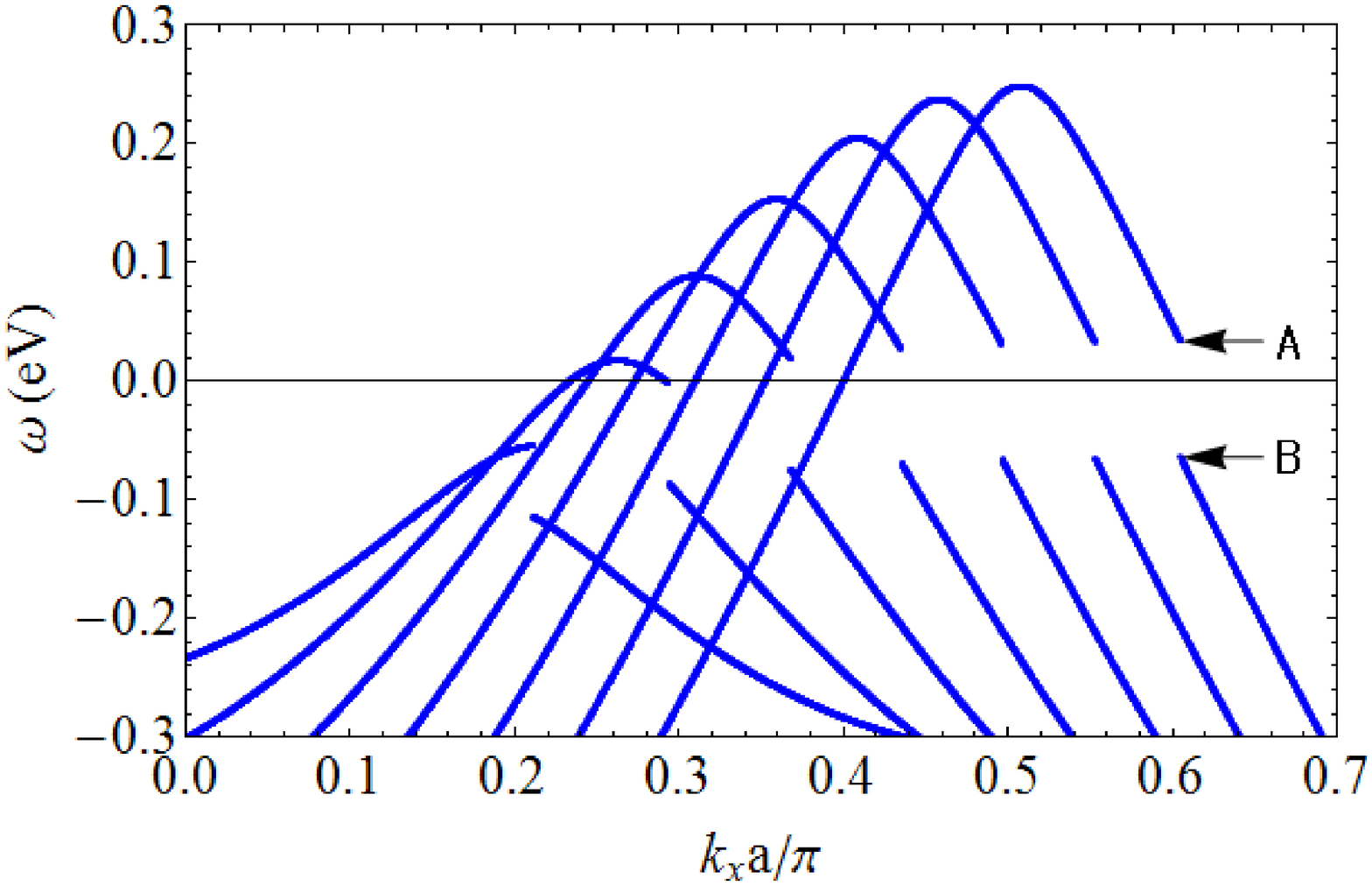}
\caption{The hole and electron band dispersions, $E_-(\vk)$ and
$E_+(\vk)$, along the cuts parallel to the nodal cut indicated
with the green dashed lines in Fig.\ 1 with the same parameters.
Fig.\ 2(a) shows the hole band in solid blue and electron band
dispersion in dashed green curve along the nodal cut. Notice that
the hole band dispersion has a gap of $2\omega_b$ at $k_x a/\pi
\approx 0.6$. The detailed hole band dispersions are plotted in
Fig.\ (b) along the cuts of $k_y a/\pi=k_x a/\pi+0.1 n$ with
$n=0-6$ from right to left. The gap survives beyond $k_y a/\pi=k_x
a/\pi+0.4$. It means that the gapless portion of the FS forms an
open ended arc as shown with the thick red solid curve in Fig.\
1.}
 \label{fig:dispersion}
\end{figure}

Fig.\ 3 is the 3D plot of the spectral function $A(\vk,\omega)$
as a function of $\vk$ at $\omega=0$. Fig.\ 3(a) is for
$\omega_b=0.05$ eV, $\alpha=0.1$ eV, and $x=12$ \%. Because of the
dynamically generated gap close to the shadow FS discussed above,
the spectral peak shows up only over a part of the FS instead of
a closed loop as in Fig.\ 3(b). Also the spectral peaks from the
electron band show up around $(0,\pi)$ and $(\pi,0)$ for a weak
$\alpha$. Now, the FS may evolve to the Fermi pocket as the
coupling $\alpha$ is increased. As an illustration, Fig.\ 3(b) is
the 3D plot of the spectral function for $\alpha=0.2$ eV with all
other parameters fixed. The Fermi pocket is clearly formed. The
peaks from the electron band are substantially reduced. Physics
behind the Fermi arc/Fermi pocket induced by the dynamic
fluctuations is quite simple: The self-energy correction given by
Eq.\ (\ref{self-energy}) dynamically generates a gap close to the
shadow FS of magnitude of about $2\omega_b$, marked by
``$2\omega_b$'' in Fig.\ 2(a). As $\alpha$ increases, the gap
between the electron and hole bands marked with ``$2\alpha$'' in
Fig.\ 2(a) becomes larger and the hole dispersion $E_-(\vk)$ of
Eq.\ (\ref{dispersion}) is pushed down. Consequently, the qp
states above the $2\omega_b$ marked with ``A'' in Fig.\ 2(b)
touch the FS. Then FS forms over a closed loop, which is the
Fermi pocket.

\begin{figure}
\includegraphics[width=5cm]{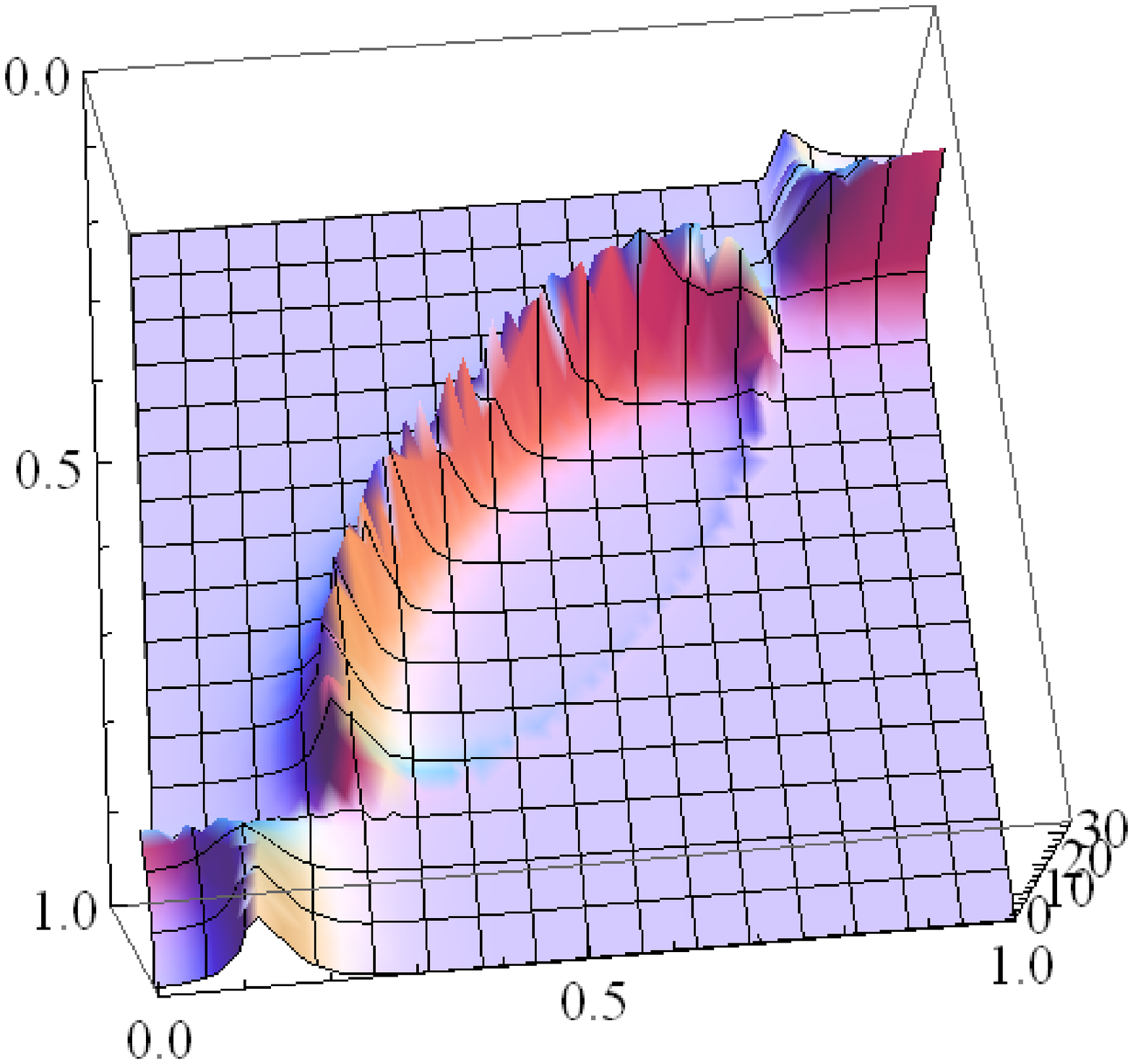}
\includegraphics[width=5cm]{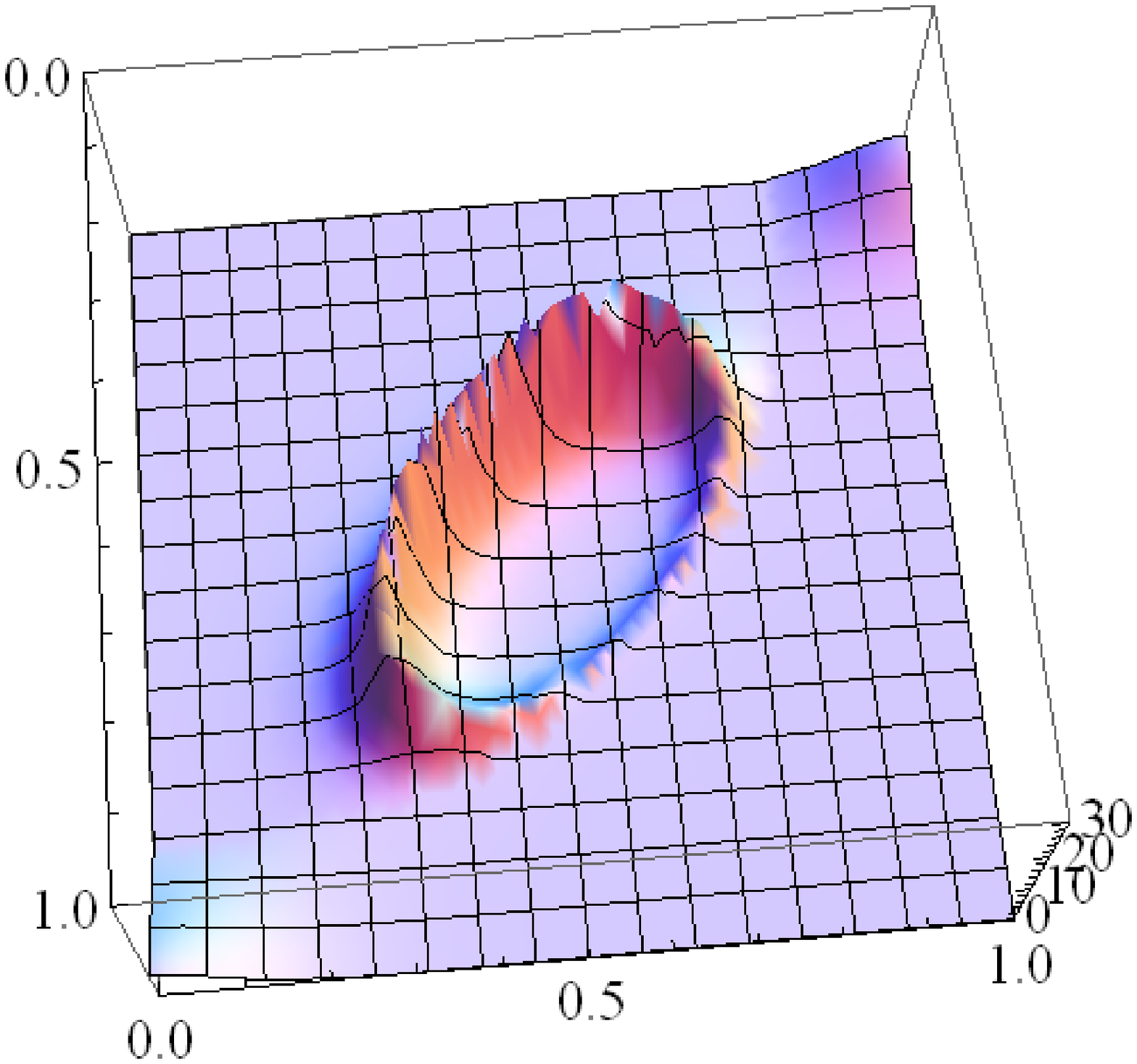}
\includegraphics[width=5cm]{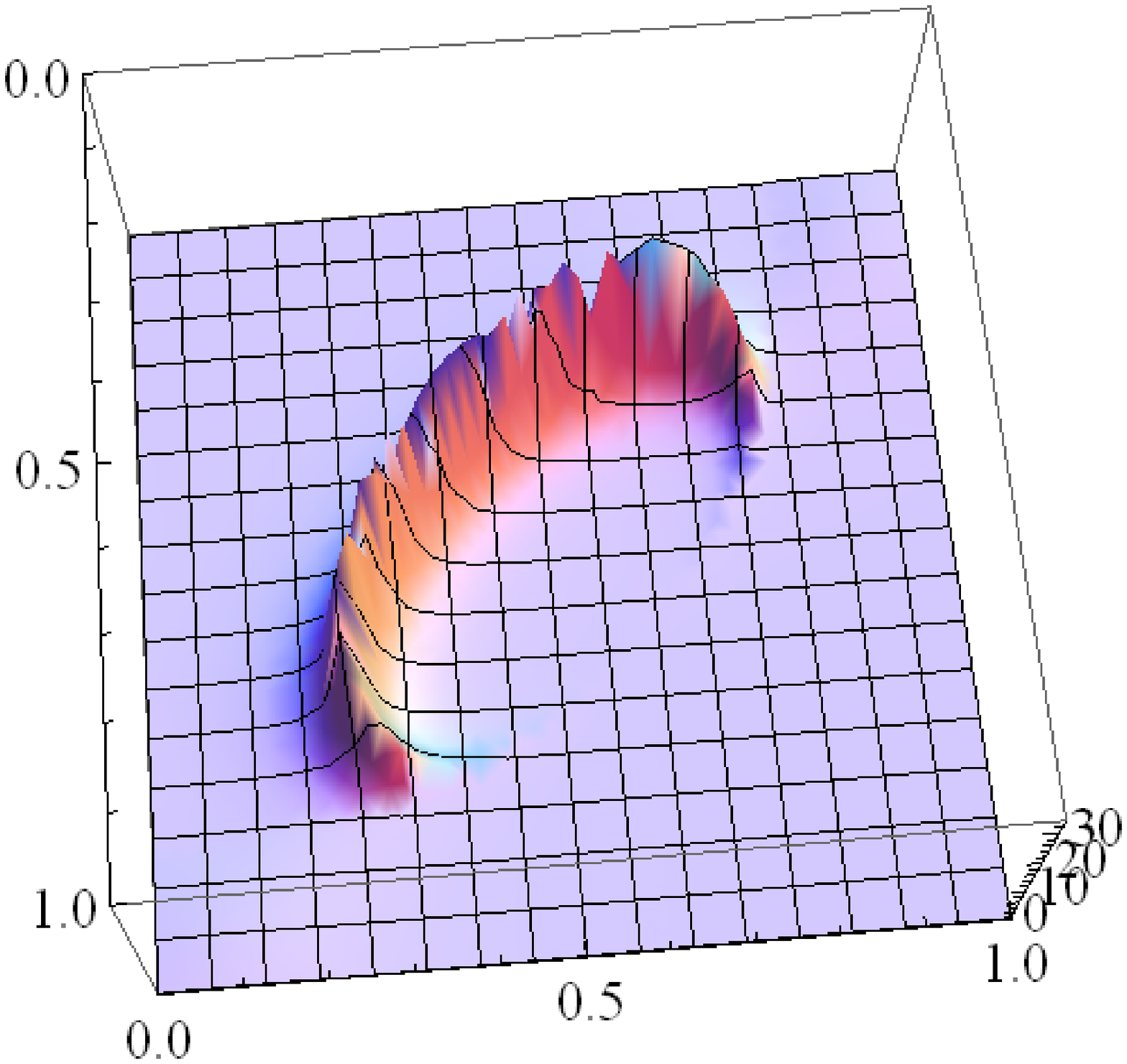}
\caption{3D plot of the spectral function $A(\vk,\omega=0)$.
Fig.\ (a) is for $\omega_b$=0.05, $\alpha=0.1$, and $\mu=-0.208$
eV corresponding to $x=12$ \%. The Fermi arc appears because of
the dynamically generated gap of magnitude of $2\omega_b$ close
to the shadow FS. For Fig.\ (b), the parameters are the same as
Fig.\ (a) except $\alpha=0.2$. The Fermi pocket appears now
because the gap is pushed down below the Fermi energy. Fig. (c)
is for $|\alpha(\vk,\vk')|^2 = \alpha_0^2 \left| \vk\times\vk'
\right|^2 $ with $\alpha_0=0.3$ and all other parameters the same
as (b). The formation of the Fermi pocket is less favored.}
\label{fig:spectral}
\end{figure}

Where $E_-(\vk)=0$ either Fermi arc or Fermi pocket shows up. If
two $\vk$ points satisfy $E_-(\vk)=0$ along any cut between the
two hot spots and parallel to the nodal cut, then the Fermi
pocket is produced. If, on the other hand, either one or two
$\vk$ points satisfy $E_-(\vk)=0$, then a portion of a pocket is
missing, which is just the Fermi arc. Both cases can be produced
with the simple formula of Eq.\ (\ref{specfun}) depending on the
parameters as discussed above.

Also interesting is the relative weights of the two peaks of the
Fermi pocket. For example, along the nodal cut, there appear two
peaks near the main band and shadow band as a function of the
momentum amplitude. The ratio of the spectral weight on the back
side of the pocket to that on the main FS is from Eq.\
(\ref{cohfactor})
 \ba
\frac{v_k^2(\xi_{k_Q}=0)}{v_k^2(\xi_{k}=0) } \approx 2\left(
\frac{\alpha} {\xi_{k_Q} } \right)_{\xi_k=0}^2 \sim 0.01
 \ea
in accord with the experimental observation.\cite{Meng09nature}

We also considered the momentum dependent coupling
$\alpha(\vk,\vk')$ as suggested by Varma and
coworkers\cite{Varma06prb,Vivek10prb} and also by Yang
$et~al$.\cite{Yang06prb}
 \ba
|\alpha(\vk,\vk')|^2 = \alpha_0^2 \left| \vk\times\vk' \right|^2.
 \ea
This form of coupling will modify the qp disperion less along the
nodal cut because $\vk\times\vk' \approx 0$ there. The Fermi
pocket formation is less favored. In Fig.\ 3(c), we show
$A(\vk,\omega=0)$ for $\alpha_0=0.3$ eV with all other parameters
the same as Fig.\ 3(b). The Fermi arcs are formed instead of the
Fermi pocket as anticipated.

From the shapes of the Fermi arcs shown in Figs.\ 1 and 3, one may
notice that the arcs turn in near the ends. It means that the
Fermi arcs seem to deviate from the underlying FS near the ends.
Norman $et~al.$ argued that this is a generic feature of the
pseudogap induced by ${\bf q} \ne 0$ order
parameters.\cite{Norman07prb} This point seems to apply to the
Fermi arcs induced by dynamic fluctuations as well although the
turning in looks weaker.

Now we understood the basic physics underlying the Fermi arc and
Fermi pocket formation with the simple dynamic bosonic
fluctuations of $\omega_b=0.05$ eV and $\xi=\infty$. But, as
$\xi\ra\infty$ the boson mode must get soft and approach
$\omega_b\ra 0$. This relation was not satisfied in the simple
case just presented. We therefore performed the full
self-consistent calculations in the following section with finite
$\xi$ and temperature. The important message of the numerical
calculations will be that the dynamically generated gap of
$2\omega_b$ in the back-side of the pocket as shown in Fig.\ 2(a)
remains intact as can be seen from Fig.\ 4(a). It means that the
qualitative feature of the FS evolution from the large FS to
Fermi arcs to Fermi pockets is unaffected. This is easy to
understand. The magnitude of discontinuity being determined by
$\omega_b$, it is insensitive to $\xi=\infty$ or not as presented
in the following section.

\section{Numerical results}

The previous discussion is based on approximate solution of the
self-energy of Eq.\ (\ref{spectral1}). Although the approximation
permits the simple and useful results discussed in the previous
section, some of the results may be an artifact of the
approximation. We therefore performed the full self-consistent
calculations via numerical iterations of the coupled equations of
Eqs.\ (\ref{self-energy}) and (\ref{spectral}). We considered the
finite correlation length ($\Gamma=\pi/\xi \ne0$ in Eq.\
(\ref{a2f})) and non-zero temperature. The more realistic
frequency dependent $\alpha^2F(\epsilon')$ as extracted by Bok
$et~al.$\cite{Bok10prb} is also considered. The important effects
of the self-consistency are that (a) the Fermi arc and Fermi
pocket coexist and (b) the center of the Fermi pocket gets
displaced towards the zone center. The fine details are
determined by the parameters like $\alpha$, $\Gamma$, and $T$.
The non-zero $\Gamma$, non-zero temperature, or the frequency
distribution of $\alpha^2 F(\epsilon')$ smear the fine structures
out.

It is interesting to note that the laser ARPES experiments
observe that the Fermi pocket coexist with the Fermi arc. The
coexistence may be understood as follows: Let us fist consider
the hole FS. The electron FS follows the same arguments. The
spectral function of Eq.\ (\ref{specfun}) indicates that the peaks
show up as a function of $\vk$ where $E_-(\vk)=0$ or $v_k^2$ is
maximum for the hole FS. The reconstructed hole FS may appear in
the region where $\xi_k>0$ and $\xi_{k+Q}>0$ around the
$(\pi/2,\pi/2)$ point. The loci of maximum $v_k^2$ can be seen
most clearly in the limit $\alpha=0$. Inspection of the coherence
factor $v_k^2$ of Eq.\ (\ref{cohfactor}) in the limit $\alpha=0$
reveals that $v_k=1$ for $\xi_{k+Q}>\xi_k$. Simultaneously,
$E_-(k)$ needs to be close to 0 as the delta function of Eq.\
(\ref{specfun}) requires. Both conditions are satisfied where
$\xi_k=0$. It is expected that peaks are produced close to the
original FS due to the $v_k^2$ factor of Eq.\ (\ref{cohfactor}).

The coexistence may also be understood as follows: The so-called
two-pole approximation of Eq.\ (\ref{spectral1}) produces two qp
branches. Next order approximation, the three-pole approximation,
is to use Eq.\ (\ref{specfun}) to the self-energy. It produces
three qp branches. Straightforward calculations reveal that, along
the nodal cut near $\omega=0$ for example, there exist one branch
close to the bare FS, and two branches almost symmetric around the
$(\pi/2,\pi/2)$ at $(\pi/2\pm\epsilon,\pi/2\pm\epsilon)$. Between
the two, the one closer to the bare FS,
$(\pi/2-\epsilon,\pi/2-\epsilon)$, merges with the branch near
the bare FS to form the main FS, and the one at
$(\pi/2+\epsilon,\pi/2+\epsilon)$ forms the back side of the
Fermi pocket. The self-consistent calculations to be presented
below maintain this feature to produce the coexisting Fermi arcs
and pockets.

Another effect of finite $\xi$ is to exhibit the dispersion kink
near $\omega\approx\ - \omega_b$. In the limit of $\xi\ra 0$, it
is simple to see that
 \ba
\Sigma(\vk,\omega) =\alpha^2 \ln \left| \frac{\omega_b
-\omega}{\omega_b +\omega}\right| .
 \ea
Then, the slope of the qp dispersion changes from
$1+2\alpha^2/\omega_b$ to 1 as $\omega$ increases past $\omega_b$.
This dispersion kink along the nodal cut was observed by many
groups and has been the focus of intense debate.

\begin{figure}
\label{fig4}
 \includegraphics[width=5cm]{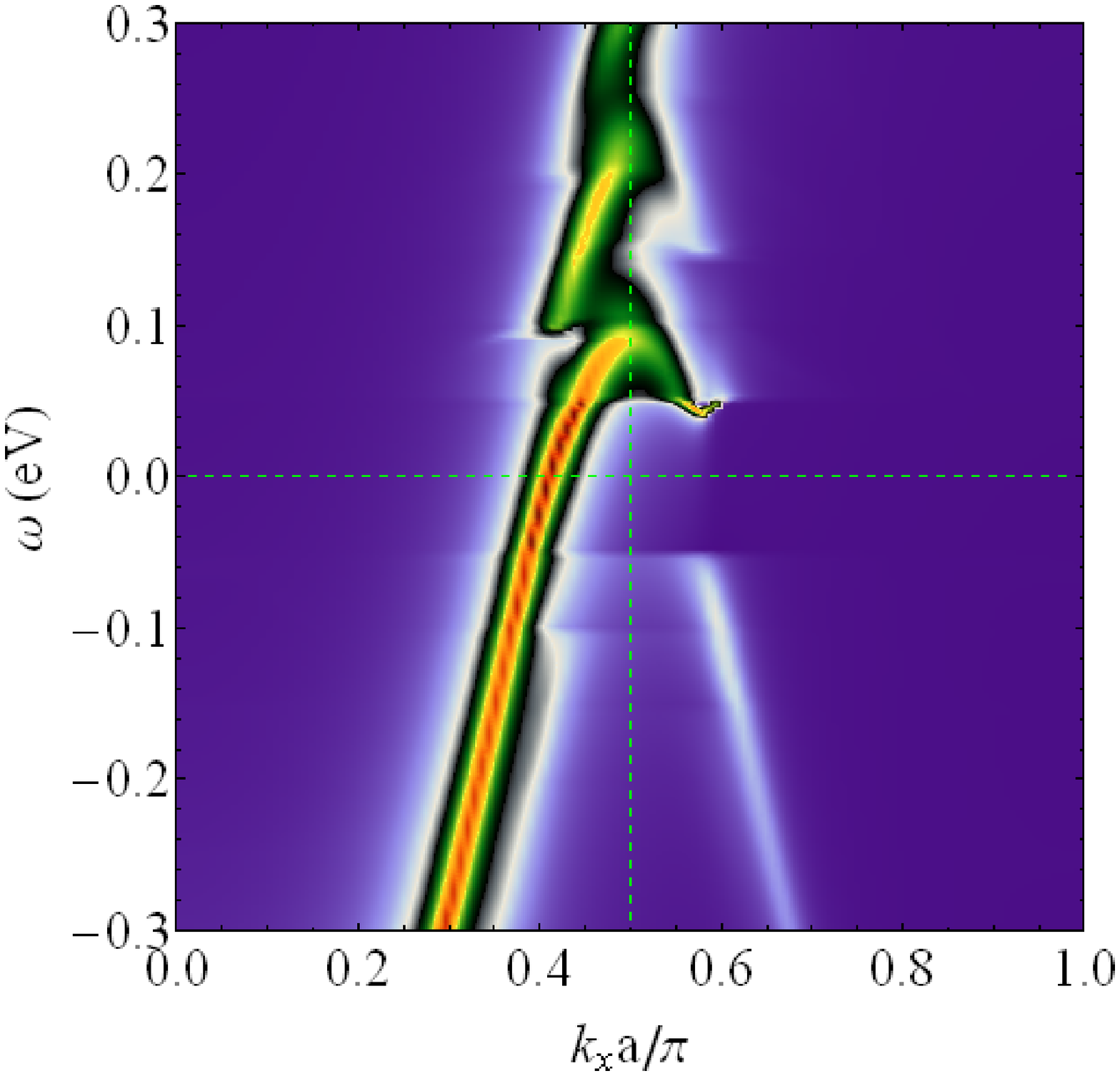}
\includegraphics[width=5cm]{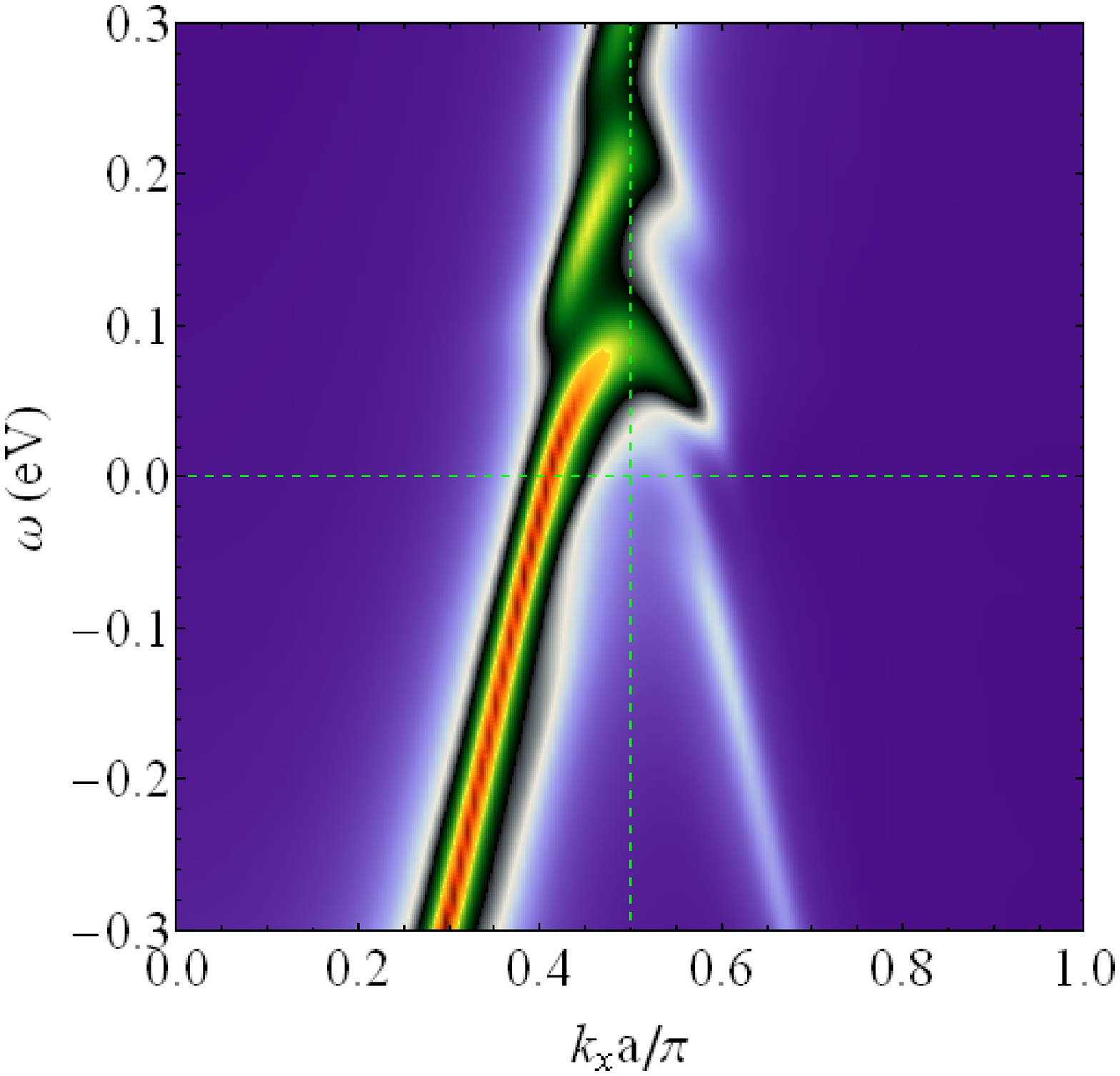}
\caption{The qp dispersion along the nodal cut from the
self-consistent calculation. $\alpha=0.18$ eV, $\Gamma=0.02$, and
$T=0$ for Fig.\ (a) and $T=200$ K for Fig.\ (b). In the ARPES
experiments the features above $\omega=0$ are cut off by the
Fermi distribution function. Notice that the gap of $2\omega_b$
opens up in the back side of the pocket for $1/\xi \ne 0$ as well.
Also notice the shadow feature around $\omega=0$ induced by the
finite temperature effects. The shadow band disperses away from
the zone center in accord with the observation by Meng $et~al.$}
\end{figure}

For finite $\xi$, the summation over $\vk'$ in Eq.\
(\ref{spectral}) is not a delta function. The $\vk'$ summation
was performed by using the 2D FFT (fast Fourier transform)
between the momentum and real spaces using the convolution
relation
 \ba
\sum_{\vk'} F(\vk'-\vk) G(\vk') = F({\bf r}) G({\bf r}).
 \ea
$2^8$ points were taken for the FFT along each axis. For $\alpha$
not too large a convergence took about 10 iterations. For $\alpha$
larger than about 0.22 eV the procedure failed to converge in our
numerical iterations. This could be an indication of a topological
change of the Fermi surface.

Fig.\ 4 is the density plot of the spectral function
$A(\vk,\omega)$ along the nodal cut as a function of $k_x a/\pi$
and $\omega$ with $\alpha=0.18$ eV, $\Gamma=0.02$, and $T=0$ for
(a) and $T=200$ K for (b). At $T=0$ the shadow band appears with
the gap of $2\omega_b$ centered around the Fermi energy. The main
band is modulated by the $\omega_b$ and the gap of $2\alpha$ is
not distinguishable. The dispersion modulation, being determined
by the energy $\omega_b$ in the case of the Einstein mode, is
expected to be weakened if the spectrum has a finite energy
distribution. This expectation is indeed the case as will be
presented below in Fig.\ 7. Also noteworthy is that the shadow
band disperses away from the $(\pi/2,\pi/2)$ as the energy is
lowered in accord with the ARPES observation of Meng $et~al.$
Compare with the lower row of the plots b--d of the Fig.\ 1 in
the Ref.\ \cite{Meng09nature}.

An important role of the finite temperature presented in Fig.\
4(b) is to bring up the qp states $below$ the Fermi energy
(marked by ``B'' in Fig.\ 2(b) for the two pole approximation) to
form the Fermi pocket. This is in contrast with the simple results
presented in the previous section. The non self-consistent
preliminary analysis indicated that the qp states $above$ the
Fermi energy (marked by ``A'' in Fig.\ 2(b)) are pushed down by
the $\alpha$ and form the Fermi pocket. This picture is modified
in the self-consistent calculations: As $\alpha$ increases the qp
dispersion above the Fermi energy bends back as can be seen from
Fig.\ 4(a) to keep the gap as intact as possible because the
total energy will be lowered by not occupying the higher lying
states. Instead the shadow band dispersion below the Fermi energy
is extended above the Fermi energy to form a pocket as can be
seen from Fig.\ 4(b). Note that at the Fermi energy the
dispersion from below is closer to the zone center than the
dispersion from above. Consequently, the pocket is displaced
towards the zone center away from the $(\pi/2,\pi/2)$ point as
shown in Fig.\ 5(a).

\begin{figure}
\label{fig5}
 \includegraphics[width=5cm]{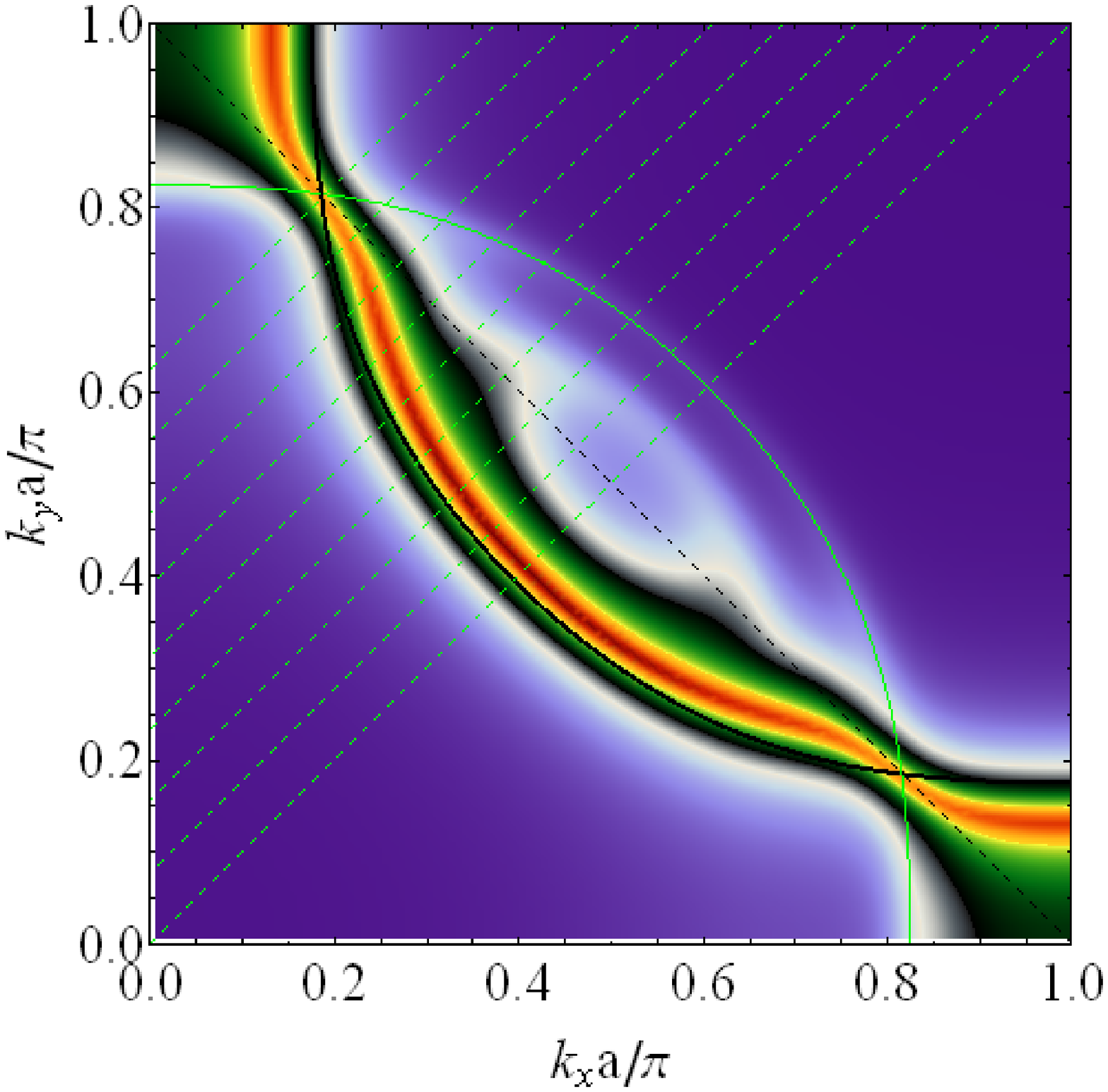}
\includegraphics[width=6cm]{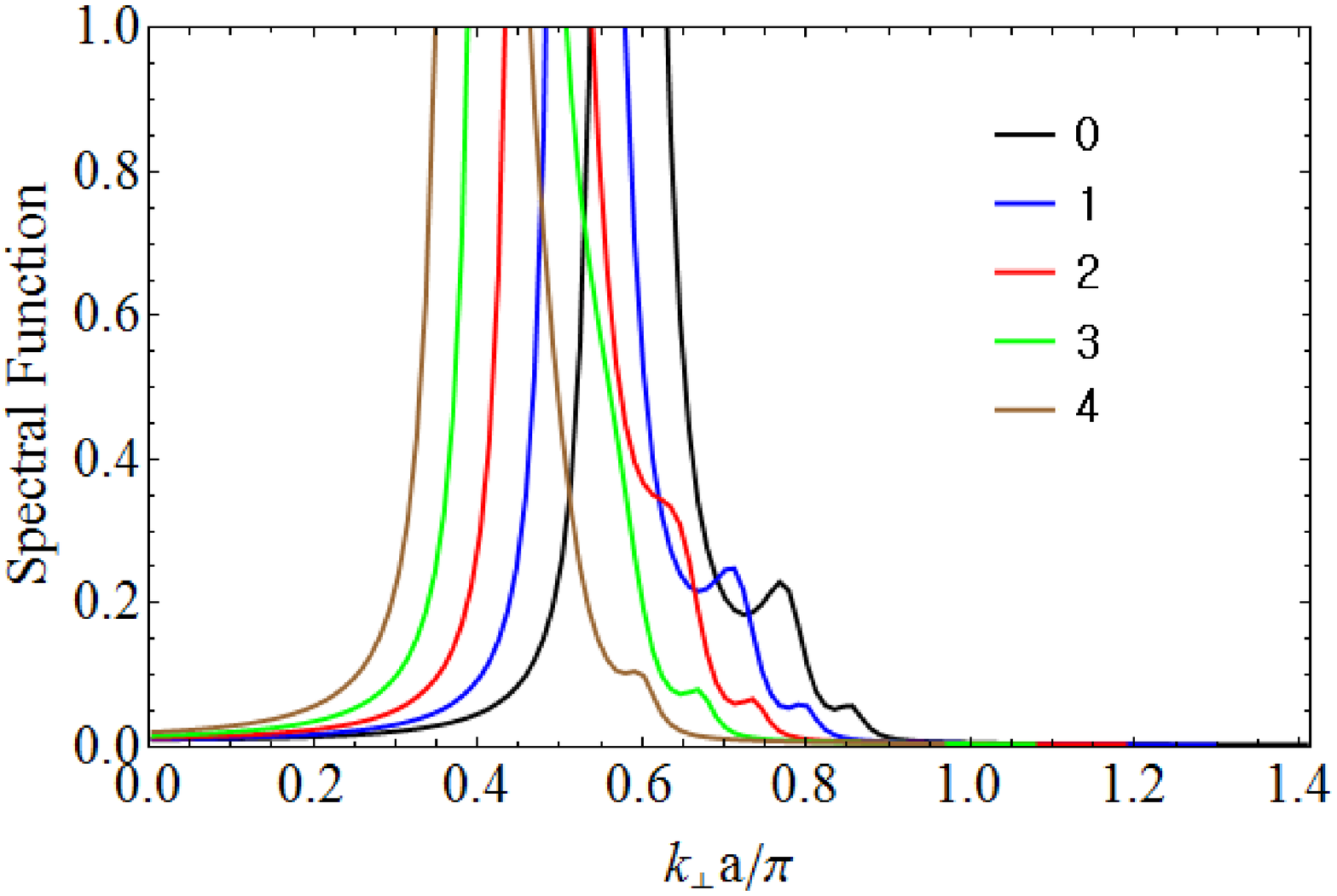}
\caption{(a) The spectral function as a function of $k_x$ and
$k_y$ at $\omega=0$ with the same parameters as the Fig.\ 4(b).
The Fermi pocket is formed because of the temperature induced
shadow feature around $\omega=0$. (b) The plots of the spectral
function of (a) along the cuts parallel to the nodal cut. From
right to left are the cuts of $k_y a/\pi= k_x a/\pi +0.2 n$ with
$n=0-4$. }
\end{figure}

In Fig.\ 5(a) we show the density plot of the spectral function
$A(\vk,\omega=0)$ as a function of $\vk$ with the same parameters
as the Fig.\ 4(b). Note the formation of the pocket coexisting
with the main Fermi surface. The center of the pocket is shifted
to the zone center away from the $(\pi/2,\pi/2)$ point as
discussed above. Fig.\ 5(b) is the plots of the spectral function
of $A(\vk,\omega=0)$ along the cuts parallel to the nodal cut.
From right to left are the cuts of $k_y a/\pi= k_x a/\pi +0.2 n$
with $n=0-4$. Note the small peaks near the back side of the
pocket. The ratio of their spectral weights to those on the main
bands is found to be about $10^{-2}$.

\begin{figure}
\label{fig6}
 \includegraphics[width=7cm]{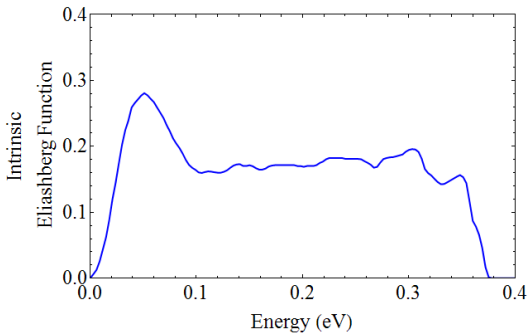}
\caption{The input Eliashberg function taken from Bok $et~al$.
The dimensionless coupling constant $\lambda\approx 1.5$. It
corresponds, for the Einstein mode of $\omega_b=0.05$ eV, to
$\alpha\approx 0.19$ eV.}
\end{figure}

We now turn to the realistic frequency dependent fluctuation
spectrum. It was taken from Bok $et~al.$\cite{Bok10prb} with a
constant $\alpha$. The input Eliashberg function $\alpha^2
F(\omega)$ is shown in Fig.\ 6. The extracted fluctuation
spectrum has a peak around $\omega\approx 0.05$ eV, flattens out
above 0.1 eV and has a cut-off at approximately 0.35 eV. The
dimensionless coupling constant
 \ba
 \label{lambdas}
\lambda=\int_0^\infty d\omega \frac{2\alpha^2F(\omega)}{\omega}
 \ea
is $\lambda\approx 1.5$. The Eliashberg function
$\alpha^2F(\theta,\omega)$ deduced by Bok $et~al.$, where
$\theta$ is the tilt angle with respect to the nodal cut and
$\omega$ is the energy, is that the functions along different
angles collapse onto a single curve below the angle dependent
cut-off energy $\omega_c(\theta)$. The cut-off is maximum along
the nodal cut, $\omega_c\approx$ 0.35--0.4 eV, and decreases as
the angle is increased. In the present calculations this angular
dependence of the cut-off energy of the Eliashberg function was
disregarded.

Eqs.\ (\ref{self-energy}), (\ref{spectral}), and (\ref{a2f}) were
solved self-consistently via iterations taking the extracted
$\alpha^2F$ of Fig.\ 6 into consideration. The $\vk'$ summation
was performed using the 2D FFT as explained above. The finite
range of the fluctuation spectrum instead of a delta function is
to smear out fine structures of the spectral function as can be
seen by comparing Fig.\ 7(a) with the Fig.\ 4 of a delta function
fluctuation spectrum. In Fig.\ 7(a) we show the dispersion along
the nodal cut at $T=100$ K, that is, the density plot of
$A(\vk,\omega)$ as a function of $k_x a /\pi$ and $\omega$. The
shadow band is also smeared out and its width is increased as the
energy is lowered. In Fig.\ (b) the density plot of
$A(\vk,\omega=0)$ is shown as a function of $\vk$. The pocket
becomes weaker compared with the delta function fluctuation
spectrum of Fig.\ 5(a).

\begin{figure}
\label{fig7}
 \includegraphics[width=6cm]{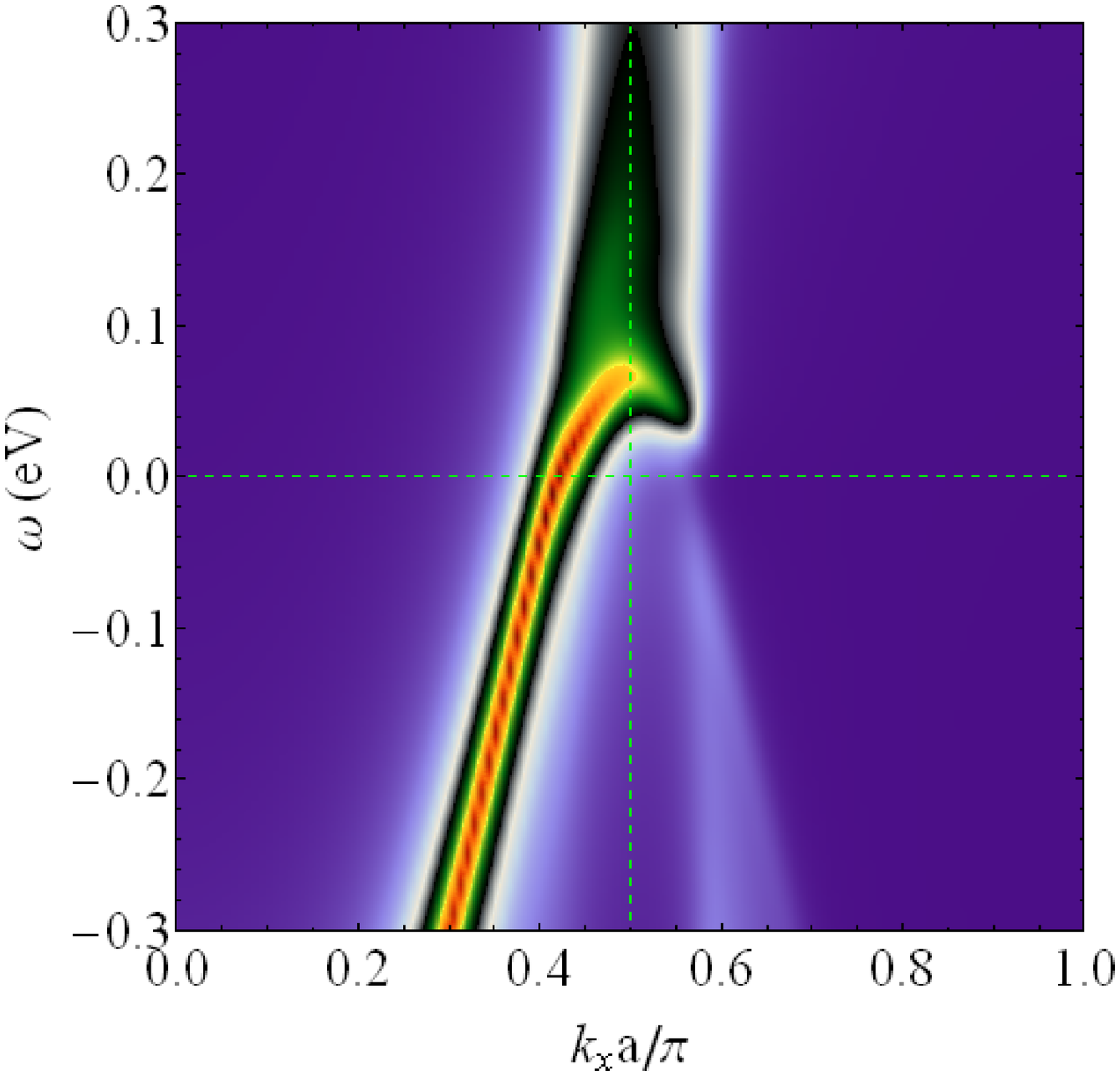}
\includegraphics[width=6cm]{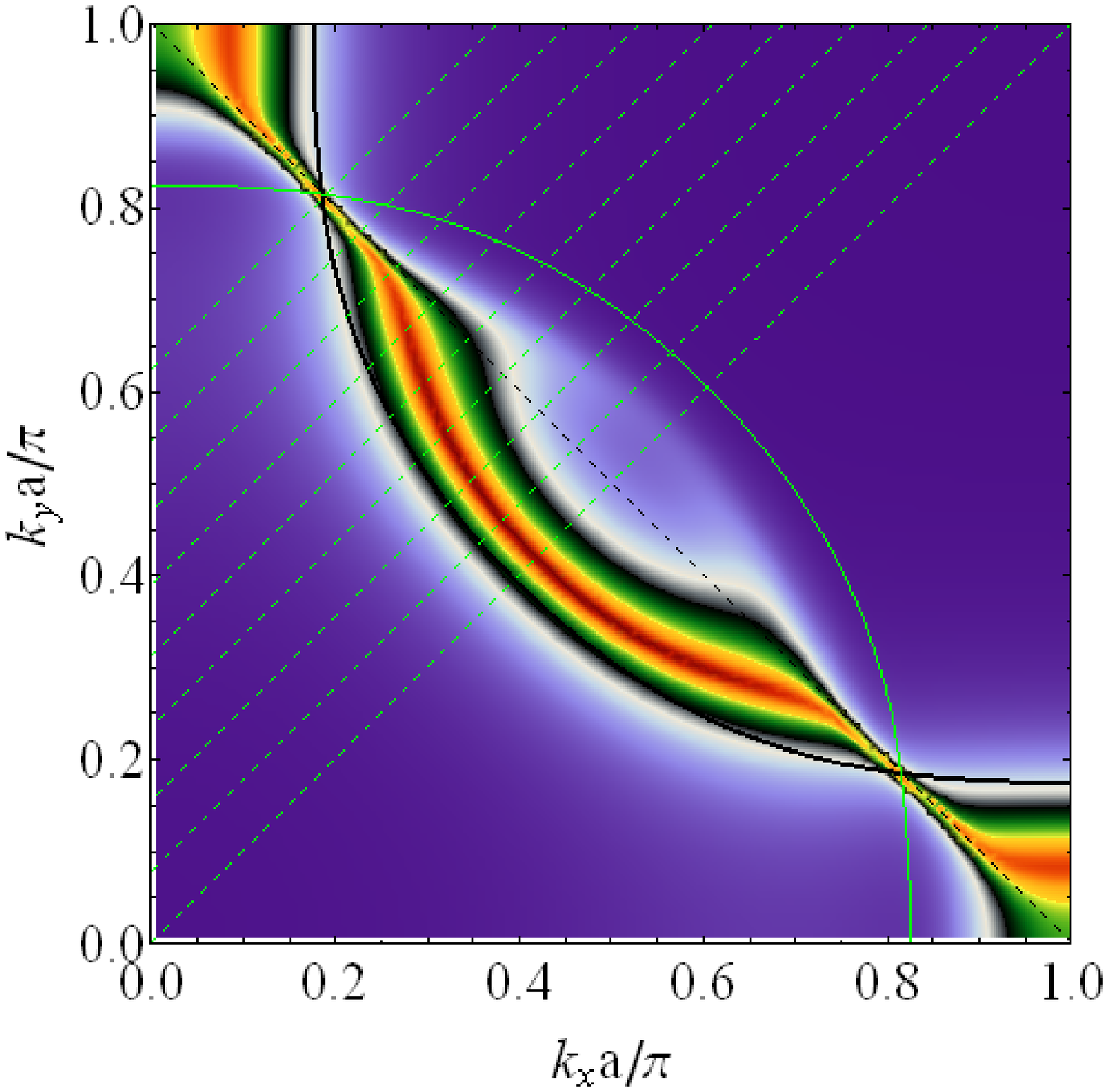}
\caption{(a) The spectral function $A(\vk,\omega)$ as a function
of $k_x a/\pi$ and $\omega$ along the nodal cuts at $T=100$ K. The
Eliashberg function of Fig.\ 6 was taken into consideration. (b)
The 3D plot of $A(\vk,\omega)$ as a function of $\vk$ at
$\omega=0$. }
\end{figure}

In order is to make a comment on implication of the coupling
constant $\lambda$ of Eq.\ (\ref{lambdas}) on superconductivity.
The approximate $T_c$ formula for $d$-wave superconductors is
 \ba
T_c = \omega_{av}\ e^{-{(1+\lambda_s)}/{\lambda_d}},
 \ea
where $\lambda_s$ and $\lambda_d$ are the coupling constant in the
normal and pairing channels, respectively. The $\lambda$ of Eq.\
(\ref{lambdas}) is $\lambda_s$ because it was extracted in the
normal state. $T_c \sim 150$ K is produced if we take $\lambda_d =
0.8 \lambda_s$. This is in accord with the expectation
$g=\lambda_d/\lambda_s <1$ for $d$-wave
superconductors.\cite{Varma10arXiv:1001.3618}

\section{Remarks and outlook}

We investigated the effects of the dynamic nature of bosonic
fluctuations on the Fermi surface reconstruction as a model for
the underdoped cuprates. The dynamic fluctuations induce the gap
of magnitude $2\omega_b$ close to the shadow Fermi surface as
Fig.\ 2 demonstrates. Then, the Fermi surface in momentum space
can be truncated unlike the Fermi surface reconstruction induced
by a long range order. Therefore, the Fermi arcs are naturally
induced by the dynamic fluctuations. The Fermi arc and/or Fermi
pocket is formed as Figs.\ 3, 5, and 7 show depending on the
coupling constant $\alpha$ or the temperature $T$ or the
correlation length $\xi$. The Fermi pocket is formed by the
filling in of the dynamically generated gap by the non-zero
temperature or the energy distribution of the bosonic spectrum
$\alpha^2 F(\omega)$. The self-consistency enables the Fermi arcs
and pockets coexist and moves their center towards the zone
center.

There have been many works along the same path adapted in this
paper, namely, employing bosonic fluctuations to compute the
renormalization of the electronic properties. See Ref.\
\cite{Varma10arXiv:1001.3618} for a recent review. Now, it will be
in order to make some comments on and comparison with a few recent
relevant works. In Ref.\ \cite{Greco09prl}, Greco computed the
electronic polarizability of $d$ density wave instability (or,
flux phase) with the $t$-$J$ model. It was used as the bosonic
fluctuations to couple to the electrons. The calculation is
non-self-consistent and assumes the true phase transition of the
flux phase. The symmetry broken phase, however, is yet to be
confirmed experimentally. Nevertheless, Greco addressed some of
the points we did not touch in this paper like the temperature
dependence of the Fermi arc length.\cite{Kanigel06naturephys} In
the absence of any symmetry broken phase in the pseudogap doping
range, however, we did not specify the mechanism of the boson
mode in this paper. Instead we took a phenomenological effective
interaction between electrons like Fig.\ 6. Because our main
point was to demonstrate the Fermi surface evolution with
$\alpha$, we did not pursue the questions like $T^*$, temperature
dependence of the arc length, and so on, leaving them as further
studies.

Dahm $et~al.$ made a check if a self-consistent description is
possible between ARPES and inelastic neutron scattering (INS) for
YBa$_2$Cu$_3$O$_{6.6}$.\cite{Dahm09naturephys} In more detail,
they fitted the INS to extract the spin susceptibility. Then they
used it as the bosonic fluctuation to couple with the electrons
to calculate the self-energy. The results were consistent with
ARPES intensity and nodal dispersion, and the kink along the
nodal cut was produced. This nodal kink is expected in their work
because $\Gamma$ of the extracted susceptibility is non-zero.

The dynamic fluctuation model with no long-range order of the
present paper successfully describes the FS evolution from the
large FS to Fermi arc to Fermi pocket as the coupling is
increased. Particularly, the enigmatic abrupt truncation of the
FS can be naturally understood. Other satisfactory features
include (a) the ratio of the spectral weight on the back side of
the pocket to that on the main side, (b) the dispersion kink in
the nodal direction around $\approx 0.05$ eV, and (c) the shadow
band disperses out as the energy is lowered below the Fermi
energy as Fig.\ 4(b) shows because the shadow feature is
``reflection'' of the main band with respect to the
$(0,\pi)-(\pi,0)$ line. In the laser ARPES experiments by Meng
$et~al.$ the shadow band was observed to disperse out as the
binding energy increases. See the lower row of the plots b--d in
the Fig.\ 1 of Ref.\ \cite{Meng09nature}.

Despite these satisfactory features of the dynamic fluctuations
there are some discrepancies compared with experimental
observations. First of all, the current scenario requires quite
long correlation length of order of $\xi/a \sim 10$ for the Fermi
arcs or Fermi pockets to appear. But, one of the present authors
recently inverted the high resolution laser ARPES from \bi \ in
pseudogap state to extract the bosonic fluctuations spectrum shown
in Fig.\ 6. It was found that the correlation length is of the
order of $\xi/a \lesssim 0.1$.\cite{Bok10prb} Although the
Eliashberg function was extracted in \bi \ and the Fermi
pocket/arc was observed in \bsl, both experiments were carried
out in the pseudogap state and this contradiction needs to be
reconciled.

Secondly, if the fluctuation spectrum of Fig.\ 6 is peaked at
$(\pi,\pi)$ with $\xi/a\sim 10$, then the transport $\lambda_{tr}
\approx 2 \lambda \approx 3$ in the nodal direction because of
the $(1-\cos\theta)$ factor from the vertex correction. To our
knowledge, this large $\lambda_{tr}$ was not observed in the
resistivity measurements. Bok $et~al.$ concluded that the
correlation length must be small, $\xi \ll a$, and the spectrum
can not be from the $(\pi,\pi)$. The enhancement of
$\lambda_{tr}$ over $\lambda$ is not expected. An interesting
point in this context though is the observation by Schachinger
and Carbotte.\cite{Schachinger08prb} They compared the $\alpha^2
F$ from infrared (IR) spectroscopy and ARPES and found that they
agree well overall (after scaling) except that $\alpha^2 F$ from
IR is larger than that from ARPES around 0.06 eV by the factor of
approximately 2. This may be understood if the peak around 0.06
eV is dominantly from $(\pi,\pi)$ and the rest of the spectrum is
momentum independent. However, this scenario seems to be at odds
with the conclusion of Bok $et~al$.

As the coupling constant $\alpha$ increases, the electron Fermi
surfaces disappear first leaving the hole Fermi surfaces only as
the two-pole approximation illustrates in Fig.\ 3. This
topological change of the Fermi surface, however, was not
obtained in the self-consistent calculations because the
iteration procedures failed to converge for $\alpha$ larger than
approximately 0.22 eV. The $\alpha\gtrsim 0.2 $ eV and/or momentum
dependent $\alpha$ is expected to give interesting results about
the Fermi surface evolution as a function of doping, coupling
constant, and temperature.

It should be also interesting to check if one can understand the
quantum oscillations under the applied magnetic field with the
current scenario. It is conceivable that the dynamically induced
hole Fermi arcs/pockets are suppressed and the electron pockets
are formed as the field is applied as the quantum oscillation
experiments imply.

Finally, we wish to note that Chang $et~al.$ also observed the
back-side of the Fermi pocket in the pseudogap state in
La$_{1.48}$Nd$_{0.4}$Sr$_{0.12}$CuO$_4$ where the orthorhombic
distortion is not the primary cause.\cite{Chang08njp} Recall that
the previous observations of the shadow bands were found to be due
to the orthorhombic structural distortion.\cite{Mans06prl} This
structural feature was separated out in Meng $et~ al.$ Also the
improved resolution of the laser ARPES facilitated their
observation of the Fermi arcs and pockets. An interesting point
is that the observed shadow band by Chang $et~al.$ was much
stronger than Meng $et~al.$ and was more symmetric with respect to
the $(\pi,0) - (0,\pi)$ line. It remains to be sorted out what
causes the differences between Meng $et~al.$ and Chang $et~al$.

\begin{acknowledgments}

This work was supported by by National Research Foundation (NRF)
of Korea through Grant No.\ NRF 2010-0010772.

\end{acknowledgments}

\bibliographystyle{apsrev}
\bibliography{ref}

\end{document}